\documentclass[twocolumn,showpacs,amsmath,amssymb,prl]{revtex4}

\usepackage{graphicx}

\begin{document}

\title{Re-entrant localization of single particle transport in disordered
Andreev wires}
\author{N. B. Kopnin $^{(1,2)}$}
\author{A. S. Mel'nikov $^{(3,4)}$}
\author{V. M. Vinokur $^{(4)}$}
\affiliation{$^{(1)}$ Low Temperature Laboratory, Helsinki
University of
Technology, P.O. Box 2200, FIN-02015 HUT, Finland,\\
$^{(2)}$ L. D. Landau Institute for Theoretical Physics, 117940
Moscow, Russia\\
 $^{(3)}$ Institute for Physics of Microstructures,
Russian Academy of Sciences, 603950, Nizhny Novgorod, GSP-105,
 Russia,\\
$^{(4)}$ Argonne National Laboratory, Argonne, Illinois 60439 }

\date{\today}

\begin{abstract}
We study effects of disorder on the low energy single particle
transport in a normal wire surrounded by a superconductor. We show
that the heat conductance includes the Andreev diffusion
decreasing with increase in the mean free path $\ell $ and the
diffusive drift produced by a small particle-hole asymmetry, which
increases with increasing $\ell$. The conductance thus has a
minimum as a function of $\ell$ which leads to a peculiar
re-entrant localization as a function of the mean free path.
\end{abstract}

\maketitle

Transport phenomena in mesoscopic wires with dimensions much less
than the dephasing length have been studied for several decades;
their mechanisms are nowadays well understood (see
\cite{imry,datta,been} for review). Depending on the ratio between
the wire length $d$ and the mean free path of electrons $\ell$ one
can separate different transport regimes. The ballistic regime
holds for small wire lengths $d\ll\ell$, here the conductance is
given by the Sharvin expression, $G=(e^2/\pi\hbar)N$, where $N\sim
(k_F a)^2$ is the number of transverse modes, $a$ is the wire
radius, $a\ll d$, and $k_F$ is the Fermi wave vector. For a
shorter $\ell\ll d$, transport is diffusion controlled with the
ohmic behavior $G\sim (e^2/\pi\hbar)N\ell/d$, neglecting the
weak-localization interference between scattered electronic waves.
With a further decrease in the ratio  $\ell /d$, the ohmic
dependence breaks down due to the localization effects: the
conductance decays exponentially \cite{Thouless} when $\ell
/d<N^{-1}$.

This textbook picture holds if the transverse confinement of the
electronic states inside the wire is caused by an insulating gap
in the material surrounding the wire, which results in elastic
scattering of electrons at the wire walls with large momentum
transfer (normal reflections). In the present Letter we consider
another realization of a normal-metal wire conductor where the
electronic states are confined  by a surrounding superconducting
material. The superconducting gap $\Delta$ outside the normal wire
suppresses to zero the density of states (DOS) of single-particle
excitations for energies $\epsilon< \Delta$, thus localizing them
in the transverse direction within the wire. These states are
essentially determined by the particle-hole Andreev reflections
with low momentum transfer at the superconducting/normal-metal
(SN) boundaries. If the normal reflection processes at the SN
boundaries can be ignored, we refer to such a normal-metal
conducting region {\it inside} a superconducting environment as to
an ``Andreev wire''. An Andreev wire can be connected through bulk
normal-metal leads to an external measuring circuit. Note that our
definition of Andreev wire differs from that used, e.g., in Ref.
\cite{Reulet} where it was applied for a normal conductor in an
insulating environment, {\it connected} to superconducting leads.
A simple way to create Andreev wires is to introduce vortex lines
in a type-II superconductor or to drive a type-I superconductor in
an intermediate filamentary state by applying a magnetic field.
Andreev wires can be manufactured artificially in the form of
normal channels in a superconducting matrix, using modern
nano-fabrication techniques employed also for producing a wider
class of hybrid SN structures such as Andreev interferometers
\cite{esteve} and billiards \cite{billi}. Experimentally, the main
distinction between Andreev wires and usual conductors is that the
measurements of thermal conductance are more appropriate to probe
the single electron transport in Andreev wires because the
single-particle part of the charge transport is short-circuited by
the supercurrent.

As shown in Ref.\cite{MV,KMV} in the ballistic limit $\ell \gg d$,
Andreev processes suppress the single electron transport for all
quasiparticle trajectories except for those which have momenta
almost parallel to the wire thus avoiding Andreev reflection at
the walls. The particles confined due to Andreev reflections do
also participate in the transport but through a slow drift along
the transverse modes (Andreev states) with the group velocity $v_g
=\hbar ^{-1}\partial \epsilon _{k_z }/\partial k_z \sim \epsilon
/p_F$ much smaller than $v_F$. This Landauer-type drift
contribution is the lower limit of conductance reached as the
contribution of freely traversing trajectories decreases with
increasing $d$. In total, the electronic heat conductance due to
these two mechanisms is much smaller than what could be derived
from the Wiedemann--Franz law using the Sharvin conductance of a
normal conductor: The effective number of conducting modes $N_{\rm
eff}=\hbar \kappa /T$ is much smaller than $(k_Fa)^2$. The
conclusion of the suppression of the single electron transport in
clean systems is in a good agreement with experiments on the heat
conductivity in type-I and type-II superconductors in the
direction of magnetic field \cite{vinen,Suter}.

In the present paper we investigate how the low energy transport
with $\epsilon \ll \Delta$ in an Andreev wire of a radius $a\gg
\xi$ is affected by a weak disorder introduced by impurity
scattering. We consider clean wires $\ell \gg a$ and neglect
inelastic processes assuming $\ell _{\epsilon} \gg \ell$. We start
our analysis from a qualitative physical picture elucidating the
main results of our work. Let us consider a quasiparticle
propagating within the wire along a trajectory that bounces from
the normal/superconducting walls at both its ends. Neglecting the
slow drift, the distributions of particles and holes are equal at
the wall due to the Andreev reflection. Without disorder induced
scattering, the distributions remain equal throughout the wire,
thus the single-particle transport associated with these
trajectories vanishes. With disorder, the distributions of
particles and holes deviate from each other by an amount
proportional to the probability of scattering, $a/\ell$,
accumulated on their way in between the two walls. In the presence
of a temperature difference at the ends of the wire, the driving
force on the trajectory is proportional to $(a/d)(T_1-T_2)$ thus
the thermal conductance becomes $\kappa_A = (T/\hbar) N_A$, where
the effective number of modes is
\begin{equation}
N_A = A (k_Fa )^2(a ^2/\ell d) \ .
\label{NA-simple}
\end{equation}
The counter-intuitive behavior of the single-particle conductance
$\kappa_A$ which {\it increases} with {\it decreasing} $\ell$ was
first predicted by Andreev \cite{Andreev64} (see also
\cite{Bezugly}). The coefficient $A$ in (\ref{NA-simple}) appears
to be a slow function of $\ell$: $A\sim \ln (\ell /a)$
\cite{Andreev64}. For such ``Andreev diffusion'', disorder with
$a\ll \ell \ll d$ stimulates the single-particle transport as
compared to that in the ballistic limit: it opens new
single-particle conducting modes blocked by Andreev reflections in
the ballistic limit. This differs from the disorder effects in
normal-metal/insulator/superconductor systems where disorder opens
two-particle tunneling processes for electrical conductance, see
\cite{been} for review. The conductance $\kappa_A$ reaches its
maximum when the mean free path decreases down to $\ell \sim a$;
it further transforms into $\kappa _D= (T/\hbar)N_D$ for a dirty
wire $\ell \ll a$  where\cite{heat/theory}
\begin{equation}
N_D\sim \nu _F a^2D/d\sim (k_Fa)^2 \ell/d  \ . \label{N-dirty}
\end{equation}
Here $\nu_F$ is the normal-state DOS at the Fermi level, and
$D=v_F\ell /3$ is the diffusion coefficient. For a large mean free
path $\ell \sim d$ the conductance $\kappa_A$ transforms into the
ballistic expression \cite{MV} $\kappa \propto d^{-2}$.

We find that disorder also modifies the single-particle transport
due to the drift along Andreev states with a group velocity $v_g$,
which results from a small non-quasiclassical particle-hole
asymmetry. The characteristic mean free path for the drift appears
to be $\ell_{\rm eff}= v_g\tau$ which is considerably shorter than
the usual mean free path $\ell$. The thermal conductance
$\kappa_L=(T/\hbar)N_L$ associated with the disorder-modified
drift is found to be {\it proportional} to the mean free path for
$\ell _{\rm eff}\ll d$
\begin{equation}
N_L \sim \nu _F a^2 D_{\rm eff} /d \sim (k_Fa )^2
\left(v_g/v_F\right)^2 (\ell /d)  \label{NL-simple}
\end{equation}
where $D_{\rm eff}= v_g\ell _{\rm eff}= v_g^2\tau $ is effective
diffusion coefficient much smaller than $D$ that appears in Eq.\
(\ref{N-dirty}). This drift saturates at the ballistic
Landauer-type expression \cite{KMV}
\begin{equation}
N _L \sim (k_Fa )^2(v_g/v_F)
 \label{conduct-Landauer}
\end{equation}
for very long $\ell $ when $\ell _{\rm eff}\gg d$.

The total heat conductance $\kappa =(T/\hbar) (N_A + N_L)$
 includes the Andreev diffusion decreasing as $\ell
^{-1}$ and the diffusive drift that increases with increasing
$\ell$.
\begin{figure}[t]
\centerline{\includegraphics[width=0.8\linewidth]{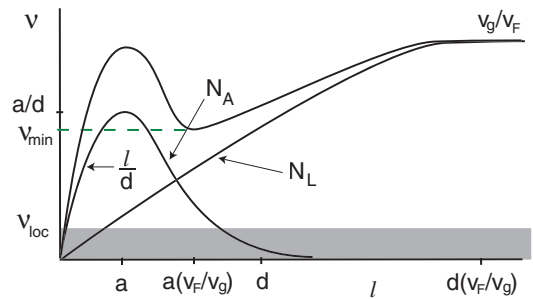}} %
\caption{Number of conducting modes (upper curve) normalized to
that in a Sharvin contact, $\nu =N/(k_Fa)^2$, as a function of the
mean free path $\ell$ interpolated according to Eqs.\
(\protect\ref{NA-simple})--(\protect\ref{conduct-Landauer}) for
the wire length $d\gg a(v_F/v_g)$. The grey area at the bottom is
the localization region, $N\sim 1$ corresponding to $\nu_{\rm loc}
\sim (k_Fa)^{-2}$.} \label{fig-localization}
\end{figure}
Equations (\ref{NA-simple}) -- (\ref{conduct-Landauer}) are
illustrated in Fig.\ \ref{fig-localization} as functions of
$\ell$. If $d\gg a(v_F/v_g)$, the number of modes has a minimum
\[
N_{min}\sim (k_F a)^2\left(a/d\right)\left(v_g/v_F\right)
\]
at $\ell _{\rm min}\sim a(v_F/v_g)$. According to the criterion of
Ref.\ \cite{Thouless}, a decrease in the number of modes down to
$N_{\rm min}\sim 1$ may lead to localization of the transport.
Thus, varying the mean free path around $\ell _{\rm min}$ we can
obtain a peculiar re-entrant localization: for a long enough
Andreev wire the quasiparticles may become localized not only in a
dirty limit $\ell \ll a$ but also for a quite long $\ell \gg a $
in such a way that the conduction opens again through either the
quasiparticle drift for longer $\ell$ or through the Andreev
diffusion for shorter $\ell$. This occurs for a wire length
$d>d_c$ where $d_c\sim a(k_Fa)^2(v_g/v_F)$.

\begin{figure}[t]
\centerline{\includegraphics[width=0.5\linewidth]{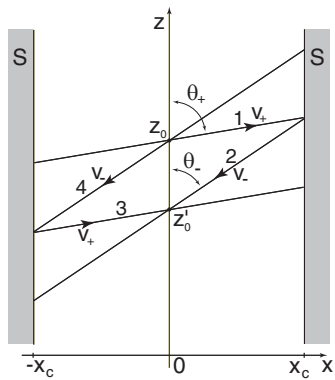}} %
\caption{Trajectories for particles (1), (3) and holes (3), (4).}
\label{fig-traject}
\end{figure}

{\it Model.\ }-- To develop a more quantitative theoretical
description we use a quasiclassical approach modified taking
account of the trajectory drift along the Andreev wire. The
excitation spectrum is \cite{Andreev65}
\begin{equation}
\epsilon _n= \frac{\hbar
v_F}{2x_c}\left(n+\frac{1}{2}\right)\sqrt{1-\frac{p_z^2}{p_F^2}} \ .
\label{spectr/slab}
\end{equation}
Here $p_z=p_F\cos \theta$ and $2x_c\sim a$ is the length of the
projection of the trajectory section between two walls onto the
plane perpendicular to the wire axis (the $z$ axis), see Fig.\
\ref{fig-traject}. Due to a $p_z$ dependence of the energy,
particles perform a slow drift along $z$ with a group velocity
\begin{equation}
v_g=\frac{\partial \epsilon _n}{\partial p_z}=-v_F\frac{\epsilon
_n }{2E _F}\frac{\cos \theta}{\sin ^2\theta} \
.\label{vgroup/slab}
\end{equation}

The group velocity can also be obtained \cite{Shytov} by
considering particle and hole trajectories, Fig.\
\ref{fig-traject}. The momentum along $x$ is $p_x^{\pm} =p_\perp
\pm m\epsilon /p_\perp $ for particles and holes, where $p_\perp
=\sqrt{p_F^2-p_z^2}=p_F\sin \theta$. These trajectories have the
direction angles $ \tan \theta _{\pm} =p_x^{\pm }/p_z$ with
respect to the $z$ axis such that $ \sin\left( \theta _+ -\theta
_-\right) =(\epsilon /\epsilon _F)\cot \theta $. The velocity
along the trajectories are
\begin{equation}
v_\pm =m^{-1}\sqrt{(p_x^{\pm })^2 +p_z^2}=v_F\pm \epsilon /p_F \ .
\label{x-velocity}
\end{equation}
Consider a particle trajectory (1) in Fig.\ \ref{fig-traject} with
the coordinates $x=s_+\sin \theta _+ , \; z=z_0+s_+\cos \theta
_+$, where $s_+$ is the distance along the trajectory. The
particle is Andreev reflected at the wall and returns back along
the hole trajectory (2) with the coordinates $x=s_-\sin \theta _-
, \; z=z_0^\prime +s_-\cos \theta _-$. Both $s_\pm $ are measured
from the wire axis. The intersection with the wall has the
coordinates $x=x_c$ and $z=z_0+x_{c}\cot \theta _+=z_0^\prime
+x_{c}\cot \theta _-$. The drift velocity defined as $v_g =v_F\sin
\theta (z_0^\prime -z_0)/2x_c$ coincides with Eq.
(\ref{vgroup/slab}). This approach is valid for wide wires $a\gg
\xi$ where the drift $z_c$ is much larger than the width
$k_F^{-1}$ of the wave packet. Since $z_c \sim a(v_g/v_F)$, this
requires $(v_g/v_F)(k_Fa)\gg 1$ which is equivalent to the
condition $\epsilon \gg \epsilon _0$, where $\epsilon _0$ is the
lowest level energy in Eq. (\ref{spectr/slab}).

{\it Kinetic equations.}-- For $\epsilon \gg \epsilon _0$ the DOS
in the normal region coincides with the DOS $\nu _F$ in the normal
state. The Boltzmann kinetic equation is
\begin{equation}
v\frac{\partial n}{\partial
s}=-\frac{1}{\tau}\left(n-\left<n\right>\right) \ .
\label{eq-Boltzmann}
\end{equation}
Here $\left< \cdots\right>$ denotes averaging over the momentum
directions. For particle and hole distributions
\[
f_+(\epsilon ,{\bf p})=n_{{\bf p}, \epsilon},\; f_-(\epsilon ,{\bf
p}) =1-n_{-{\bf p},-\epsilon}\ ,
\]
respectively, the Boltzmann equation takes the form
\begin{equation}
\pm v_\pm \frac{\partial f_\pm}{\partial
s}=-\frac{1}{\tau}\left(f_\pm -\left<f_\pm \right>\right) \ .
\label{kineq+}
\end{equation}
In both the upper-sign and lower-sign equations, the distance $s$
is measured in the direction of $+{\bf p}$.

Since all particles are Andreev reflected as holes, the
single-particle current through the wire side walls vanishes.
Using $ mv_\pm \sin \theta _\pm =p_x^{\pm }$ we put at the walls
\begin{equation}
v_+\sin \theta _+ f_+ =v_-\sin \theta _- f_- \ . \label{boundcond}
\end{equation}

The kinetic equation (\ref{kineq+}) on the trajectory (1) gives
\begin{eqnarray}
f_+(s_+) &=&f_+(0)e^{-s_+/\ell _+} \nonumber \\
&+&\ell_+^{-1}e^{-s_+/\ell _+}\int _0^{s_+}
\left<f_+\left(s_+^\prime \right)\right>e^{s_+^\prime /\ell _+}\,
ds_+^\prime \ . \quad \label{f+(s)}
\end{eqnarray}
The function $f_+(0)$ is taken at $x=0,\, z=z_0$. To get the
corresponding expression on trajectory (2) one substitutes $f_+$,
$s_+$, $\ell _+$, and $z_0$ with $f_-$, $s_-$, $\ell _-$, and
$z_0^\prime$, respectively. Putting $s_+=x_{c}/\sin \theta _+$ and
$s_-=x_{c}/\sin \theta _-$ for $f_+$ and $f_-$, respectively, we
insert the result into the boundary condition Eq.\
(\ref{boundcond}). The distributions at the trajectories (3) and
(4) for $s<0$ are found by replacing $z_0 \leftrightarrow
z_0^\prime$. The boundary condition for them is applied at
$s_+=-x_{c}/\sin \theta _+$ for $f_+$ and at $s_-=-x_{c}/\sin
\theta _-$ for $f_-$.

Assuming that $\left<f (x,z)\right>$ depends only on $z$,
\[
\left<f (s)\right>=\left<f (0)\right>+s\cos \theta \frac{\partial
\left<f \right> }{\partial z} \ ,
\]
the next step is to expand the boundary conditions in small
$v_g/v_F$ and $s_c/\ell$ for angles $\theta \gg a/\ell$.
Neglecting small terms $\partial f_2 /\partial z$ we obtain at the
wire axis $x=0$
\begin{equation}
f_2=\frac{v_g}{v_F \cos \theta} f_1 -\cos \theta
\frac{s_c^2}{2\ell} \frac{\partial \left< f_1\right>}{\partial z}\
, \label{f2/largetheta}
\end{equation}
\begin{equation}
\ell\frac{v_g}{v_F}\, \frac{\partial f_1}{\partial z} = -\left(
f_1-\left< f_1\right>\right) \ . \label{f1/largetheta}
\end{equation}
We denote $s_c = x_c/\sin \theta$ and introduce
\begin{equation}
f_1=-(f_+ +f_-), \; f_2 = -(f_+ - f_-) \label{fpm-def}
\end{equation}
in accordance with the definitions used in the theory of
superconductivity \cite{book}. The functions $f_1$ and $f_2$ are
nearly constant along the trajectory within the wire. Analysis of
Eq.\ (\ref{f+(s)}) shows that for angles $\theta \ll a/\ell$, the
distinction between the usual and the Andreev diffusion
disappears, and $ f_2=- \ell \cos \theta  (\partial \left<
f_1\right>/\partial z)$ while the counterpart of the first term in
Eq.\ (\ref{f2/largetheta}) proportional to $(v_g/v_F)f_1$
decreases exponentially.

The first term in Eq.\ (\ref{f2/largetheta}) describes the drift
along the Andreev states with the velocity $v_g$. The second term
is the Andreev diffusion \cite{Andreev64}. Equation
(\ref{f1/largetheta}) introduces an effective mean free path $\ell
_{\rm eff}= (v_g/v_F)\ell =v_g\tau$ much shorter than the usual
$\ell$. In the ballistic limit of very long $\ell$ such that
$\ell_{\rm eff} \gg d$, the distribution $f_1$ is constant along
the wire and
\begin{equation}
f_2=\frac{v_g}{v_F \cos \theta} f_1 \ . \label{f2/ballistic}
\end{equation}

In the most practical limit when $\ell _{\rm eff} \ll d$,
\begin{equation}
f_1=\left< f_1\right> -v_g\tau \frac{\partial \left<
f_1\right>}{\partial z} \label{f1/diffus}
\end{equation}
and
\begin{equation}
f_2 = -\left[ \frac{v_g^2\ell }{v_F^2 \cos \theta}+\cos \theta
\frac{s_c^2}{2\ell} \right] \frac{\partial \left<
f_1\right>}{\partial z} +\frac{v_g}{v_F \cos \theta} \left<
f_1\right>  .  \label{f2/diffus}
\end{equation}
The first term in brackets describes the disorder-modified drift.
Its relative magnitude with respect to the Andreev diffusion (the
second term) is of the order of $(v_g/v_F)^2(\ell /s_c)^2$, i.e.,
much larger than non-quasiclassical corrections of the order
$(\epsilon /E_F)^2$ to the usual diffusion. We neglect those
corrections in what follows.

{\it The energy current\ } has the form \cite{book}
\begin{equation}
I_{\cal E} = -\nu _F v_F\int d^2 r \int \frac{d \Omega _{\bf
p}}{4\pi} \int _{-\infty}^{+\infty}  \cos \theta \, \epsilon
\,f_2\, d\epsilon \ . \label{energycurrent}
\end{equation}
For very long mean free path $\ell _{\rm eff}\gg d$ the
distribution is determined by Eq. (\ref{f2/ballistic}), and we
recover the Landauer formula derived in Ref. \cite{KMV}
\begin{equation}
I_{\cal E} = -\nu _F \int d^2 r \int \frac{d \Omega _{\bf
p}}{4\pi} \int _{-\infty}^{+\infty}  \epsilon v_g\,f_1\, d\epsilon
\ . \label{I-Land}
\end{equation}
With $f_1 =\tanh(\epsilon /2T_1)$ for $v_g>0$ and $f_1
=\tanh(\epsilon /2T_2)$ for $v_g<0$ we arrive at Eq.
(\ref{conduct-Landauer}) for the heat conductance.

Assuming that the group velocity is independent of $\theta$ we
obtain a qualitative behavior described in introduction. Indeed,
for $\ell _{\rm eff}\ll d$, the distribution obeys Eq.\
(\ref{f2/diffus}) where the last term does not contribute to the
current. The current becomes $I_{\cal E} =\kappa (T_1-T_2)$ with
the total thermal conductance $\kappa =(T/\hbar)(N_A+N_L)$ given
by Eqs.\ (\ref{NA-simple}) and (\ref{NL-simple}). This simplified
picture has to be modified, however, due to a rapid divergence of
$v_g$ at small angles. This leads to a more complicated behavior
of the drift contribution to the heat conduction as a function of
temperature and of the mean free path characterized by different
power laws in different regions of $\ell$ and $T$. The detailed
analysis of the heat conduction will be published elsewhere. Here
we discuss its behavior for a long wire with $d\gg a\sqrt{E_F/T}$.
For relatively short mean free path $a\ll \ell \ll a\sqrt{E_F/T}$,
the angular integral of the first term in the brackets in Eq.\
(\ref{f2/diffus}) is cut off by the exponential decay of the drift
term for $\theta \ll a/\ell$. Therefore,
\[
N_L\sim (k_Fa)^2(T/E_F)^2(\ell ^3/a^2d) \ .
\]
This behavior is replaced with $N_L\sim (k_Fa)^2(T/E_F)(\ell /d)$
for $a\sqrt{E_F/T}\ll \ell \ll d$ and further transforms into
\[
N_L \sim (k_Fa)^2(T/E_F)L
\]
where $L\sim \ln(\ell /d)$ for $d\ll \ell \ll d(E_F/T)$ and $L\sim
\ln (E_F/T)$ for $\ell \gg d(E_F/T)$. Therefore, the full
saturation at the $\ell$-independent Landauer-type drift
conduction Eq.\ (\ref{I-Land}) occurs when $\ell _{\rm eff}\sim
(T/E_F)\ell $ becomes larger than $d$. The total conduction
reaches its minimum
\[
N_{\rm min}\sim (k_Fa)^2(a/d)\sqrt{T/E_F}
\]
at $\ell _{\rm min}\sim a\sqrt{E_F/T}$. The re-entrant
localization is possible for wire lengths longer than $d_c\sim
a(k_Fa)^2\sqrt{T/E_F}$.

To summarize, we develop a theory of single electron transport and
re-entrant localization in clean Andreev wires. Our results could
stimulate experimental research of these phenomena in the mixed or
intermediate state, as well as in hybrid SN structures.

This work was supported in part by the US DOE Office of Science
under contract No. W-31-109-ENG-38, by Russian Foundation for
Basic Research, the Program ``Quantum Macrophysics'' of the
Russian Academy of Sciences, Russian State Fellowship for young
doctors of science, and NATO Collaborative Linkage Grant No.
PST.CLG.978122. A.S.M. is grateful to the Low Temperature
Laboratory at the Helsinki University of Technology for
hospitality.

\end{document}